\documentclass[prl,twocolumn,aps,showpacs,letterpaper]{revtex4-1}
\usepackage{graphicx}
\usepackage{epsfig}
\usepackage{natbib}

\begin{document}

\title{Bang-bang shortcut to adiabaticity in trapped ion quantum simulators}

\author{S.~Balasubramanian}
\affiliation{Department of Physics, Massachusetts Institute of Technology, Cambridge, Massachusetts 02139, USA}
\author{Shuyang Han}
\affiliation{Department of Physics, Georgetown University, Washington, DC 20057, USA}
\author{B.~T.~Yoshimura}
\affiliation{Department of Physics, Georgetown University, Washington, DC 20057, USA}
\author{J.~K.~Freericks}
\affiliation{Department of Physics, Georgetown University, Washington, DC 20057, USA}

\begin{abstract}
We model the bang-bang optimization protocol as a shortcut to adiabaticity in the ground-state preparation
of an ion-trap-based quantum simulator. Compared to a locally adiabatic evolution, the bang-bang protocol produces a somewhat lower ground-state probability, but its implementation is so much simpler than the locally adiabatic
approach, that it remains an excellent choice to use for maximizing ground-state preparation in systems that cannot be solved with conventional computers. We describe how one can optimize the shortcut and provide specific details for how it can be implemented with current 
ion-trap-based quantum simulators.
\end{abstract}

\date{\today}
\maketitle

\noindent
{\it Introduction.} There has been much recent progress in ion-trap-based quantum simulation. Original experiments focused on adiabatic state preparation~\cite{kim,rajibul1,edwards} of the transverse-field Ising model by initially orienting all of the spins along the field axis (in a large initial field) and then ramping the field to zero to create the ground state of the Ising model. But when the system size was increased, and frustrated antiferromagnetic systems were examined, it became clear that these experiments would have a large amount of diabatic excitation~\cite{rajibul2}, which led to the study of excited states~\cite{rajibul2,senko1,yoshimura1,roos} and to a protocol that optimizes the field ramp with a locally adiabatic criterion~\cite{richerme1}. In addition, other experimental situations were examined, such as Lieb-Robinson bounds~\cite{senko2,blatt} and higher-spin cases~\cite{richerme2}. Currently, there are two foci for adiabatic state preparation: (i) find shortcuts which will allow the original protocol to be achieved or (ii) use the diabatic excitations as a means to study low-energy excitations. Within the first category, recent work has found an exact shortcut for adiabatic state preparation~\cite{shortcut1,shortcut2} (at least for the nearest-neighbor transverse field Ising model), but the multiple-spin interactions needed to accomplish this goal are too complicated to implement in the current generation of quantum simulators. In the second category, we already mentioned experimental~\cite{rajibul2,senko1,roos} and theoretical~\cite{yoshimura1} methods to produce or measure specific excitations. It also is possible, in some cases, for the diabatic excitations to resemble an equilibrium thermal state, especially for ferromagnetic systems~\cite{lim}.

\begin{figure}[thb]
\vskip -0.1in
\centering{
\includegraphics[width=0.75\columnwidth]{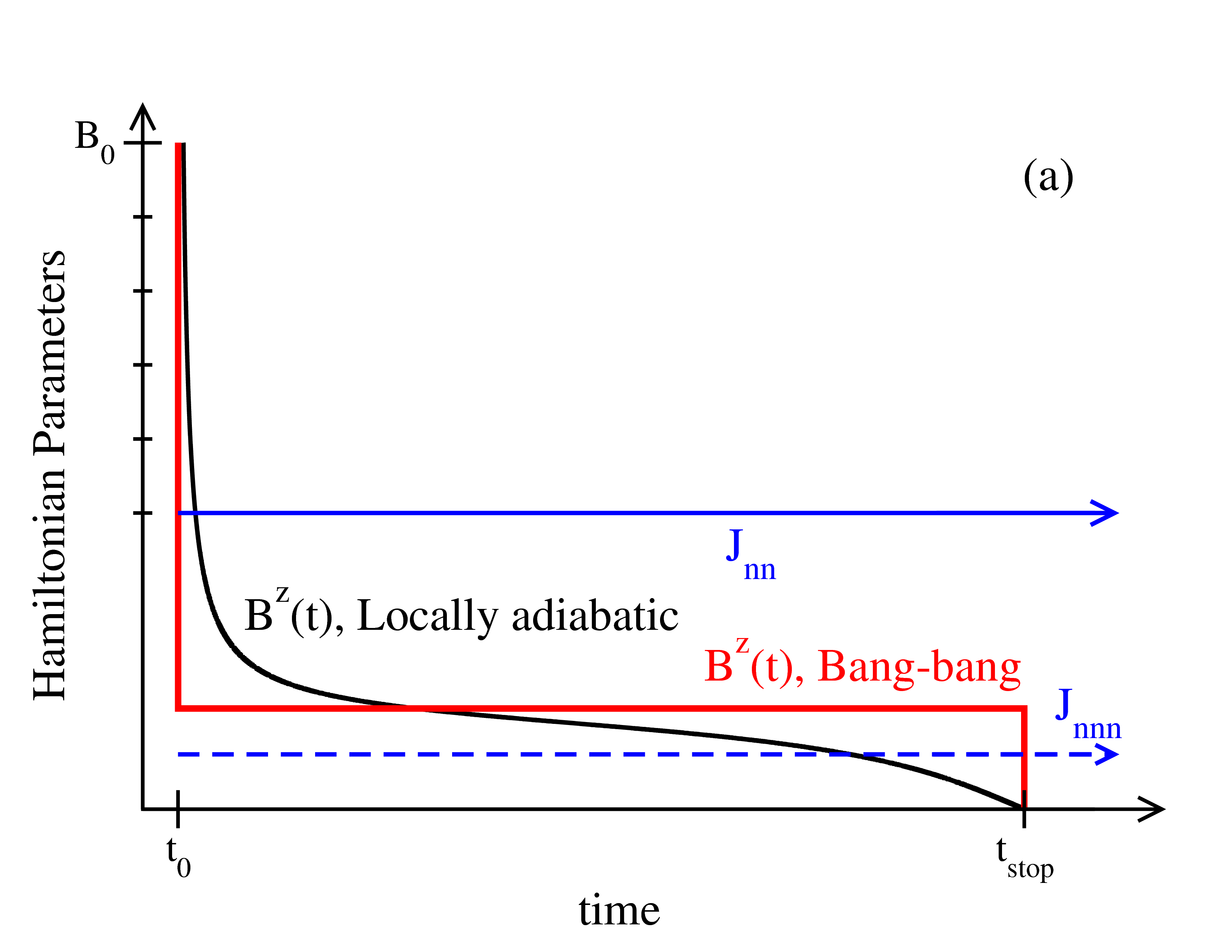}
}
\vskip -0.06 in
\centering{
\includegraphics[width=0.75\columnwidth]{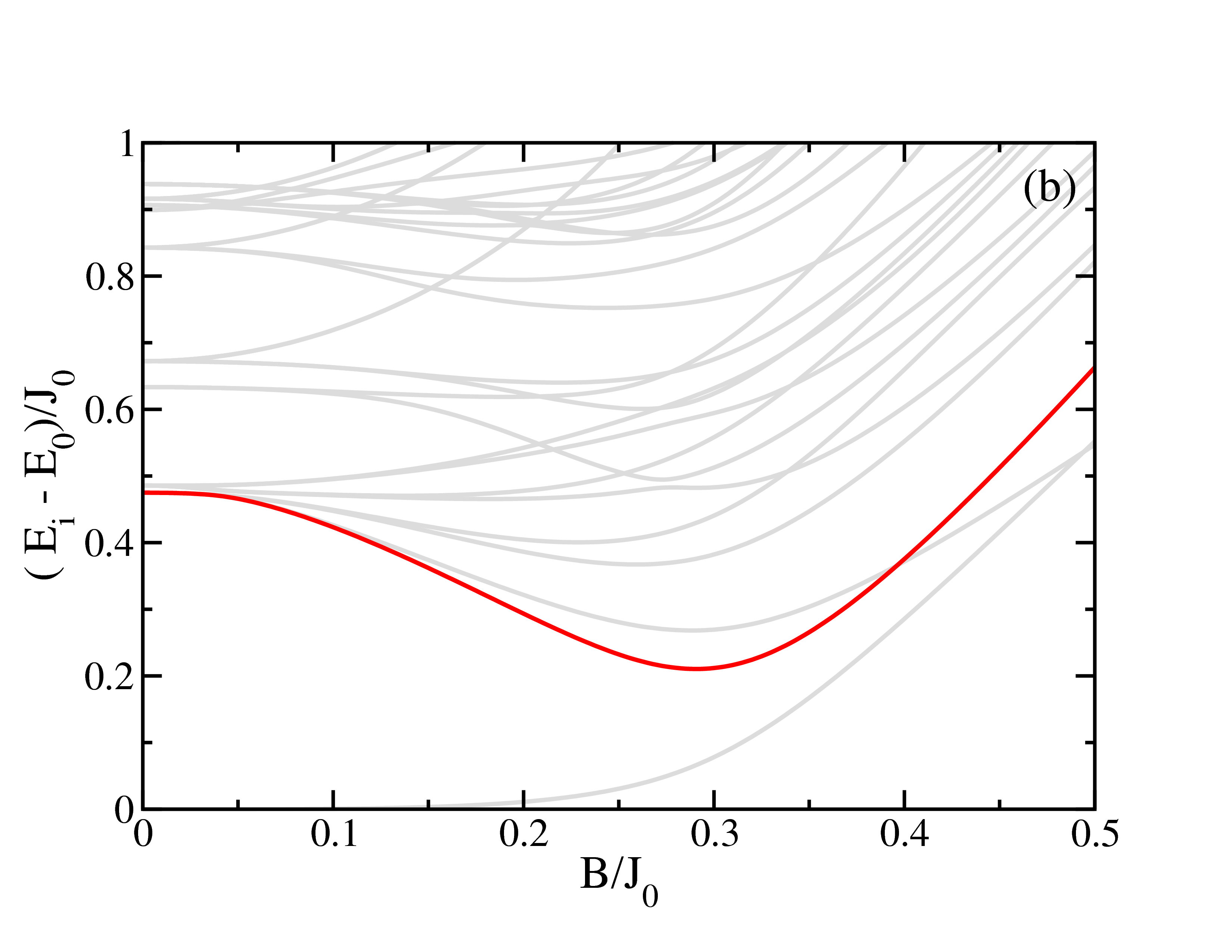}
}
\caption{(Color online.) 
\label{fig: protocol}
(a) Comparison of the different ramp protocols. The magnetic interactions are held fixed, while the field is ramped from
an initial value about five times the average nearest-neighbor exchange to zero. Two ramp profiles are shown---the locally adiabatic ramp, which strives to have a uniform rate of diabatic excitation throughout the ramp, and the bang-bang protocol. (b) The low-lying spectra of an $N=10$ Ising chain with spin-spin interactions for the trap used in Ref.~\onlinecite{rajibul2} versus the ground-state energy as a function of the magnetic field (here, the exchange coefficients decay with an approximate power law of $\alpha=1.05$). Since the transverse-field Ising model has both spatial parity and spin-reflection parity, the red line denotes the lowest-energy state that is coupled to the ground state and hence plots $\Delta(B)$, as described in the text.
}
\end{figure}

The bang-bang protocol has long been known within the field of quantum control as a useful optimization algorithm~\cite{quantum_control}. It invokes a control strategy similar to the algorithm employed with a thermostat, which sequentially turns the climate control system fully on or fully off to maintain the temperature within a specified range.  Here, it corresponds to quenching the magnetic field to an initial value, holding it for a fixed time, and then subsequently quenching it to zero. This protocol is illustrated schematically in Fig.~\ref{fig: protocol}~(a), along with the more conventional locally adiabatic ramp~\cite{richerme1}. The locally adiabatic ramp is determined by having uniformity in the diabatic excitations throughout the ramp. It strives to ramp the field quickly when the energy gap to the lowest coupled excited state is high, and more slowly when that gap is small, but it requires detailed knowledge of the energy of the first coupled excited state as
a function of magnetic field in order to determine the ramp. While this can be found experimentally utilizing different methods~\cite{senko1,roos,yoshimura1}, it is a difficult procedure to carry out for large systems that have significant frustration.
The bang-bang protocol is much simpler. It is motivated, in part, from a mathematical proof which says the most adiabatic ramp starts 
and ends with the flattest field profile~\cite{lidar}; the bang-bang approach carries this functional form to its extreme limit. But because we found that the optimal ramp always was the locally adiabatic ramp, the mathematical proof must not hold for this
class of experimental ramps.
The bang-bang shortcut works by projecting the initial state onto a collection of eigenstates at an intermediate field, allowing 
those states to evolve in the constant field until the projection onto the field-free ground state is maximized, when the quench to zero  field occurs (also done by projection onto the field-free eigenstates). It is not clear whether waiting longer times will necessarily improve the bang-bang shortcut, although our results certainly suggest that improvements in the
final ground-state probability do occur if one runs the experiment over a longer period of time. It is likely that this is intimately related to quantum speed limits~\cite{quantum_speed_limits}.

In particular, we found that the bang-bang shortcut produces about 80\% of the ground-state probability that the locally adiabatic 
protocol produces, when both experiments are run over about the same period of time. Nevertheless, because of its simplicity in implementation, it remains an attractive alternative temporal profile for the field ramp. In addition, it provides a different perspective for understanding diabatic excitations within quantum simulators.

\noindent
{\it Formalism.} The Hamiltonian for the transverse-field Ising model is
\begin{equation}
\mathcal{H}(t)= -\sum_{\scriptsize\begin{array}{c}
i,j=1\\
i<j
\end{array}}^N J_{ij}\sigma^x_{i}\sigma^x_{j}-B^{z}(t)\sum_{i=1}^N\sigma^z_{i}.
\label{eq: ham}
\end{equation}
Here, $\sigma^r_{i}$ is the Pauli spin matrix (with eigenvalues $\pm 1$ and with $r=x$, $y$, or $z$ denoting the spatial direction of the Pauli matrix) at lattice site $i$, $B^z(t)$ is
the time-dependent transverse field, and $N$ is the number of spins in the lattice; we work in units with $\hbar=1$
and simulate the transverse-field Ising model in a linear Paul trap.
Experimentally, the model is generated by using clock states of the $^{171}$Yb$^{+}$ ion as the spin up and spin down states and then driving the system with a laser-induced spin-dependent force. This is achieved by employing both red and blue detuned laser beams from the carrier transition which induce a $\sigma^x$ operation on the hyperfine states, whose strength is proportional to the phonon coordinate at lattice site $i$. Integrating out the phonons, under the assumption that they are only virtually occupied during the experiment, yields the following static spin exchange coefficients---after averaging over their time dependence---\cite{duan_monroe} (we use conventional frequency units for all parameters):
\begin{equation}
J_{ij}=\Omega^2\nu_R\sum_{m=1}^N\frac{b_{i}^{m}b_{j}^{m}}{\mu^2-\omega_m^2}.
\end{equation}
We use the experimental parameters from Ref.~\onlinecite{rajibul2} where $\Omega= 600$~kHz is the Rabi frequency, $\nu_R=h/(M\lambda^2)=18.5$~kHz is the recoil energy of a $^{171}{\rm Yb}^+$ ion (with $h$ being Planck's constant, $M$ the mass of the ion, and $\lambda=355$~nm the wavelength of the laser light). In addition, $b_{i}^{m}$ is the value of the orthonormal eigenvector at the $i^{\rm th}$ ion site of the $m^{\rm th}$ transverse normal mode for the $N$-ion chain, $\omega_m$ is the corresponding normal mode frequency, and $\mu$ is the detuning of the laser from the transverse center of mass mode.   We let $J_0<0$ denote the average nearest-neighbor spin-spin interaction for the antiferromagnetic case. The axial center of mass mode has its frequency adjusted from 620~kHz to 950~kHz, corresponding to a nearest neighbor exchange interaction which is near 1~kHz ($|J_0|\approx 1$~kHz); the exchange coefficients decay with an approximate power law that ranges from $0.7<\alpha<1.2$. 

For the bang-bang optimization, the time-evolution of the wavefunction is trivial to calculate. Each quench is handled by the sudden approximation, where one takes the overlap of the current state with the eigenstates of the quenched Hamiltonian. The time evolution for the intermediate fixed-field Hamiltonian is also trivial, since the Hamiltonian is time independent, so the eigenstates evolve with a linearly increasing phase determined by their eigenvalues and the time of the evolution. Both the quench field and the hold time are varied to optimize the final ground-state probability. An experimental implementation requires determining the probability to be in the final ground state to carry out the optimization. This might be difficult to achieve if the system is so complex that one does not know {\it a priori} what the ground state is, but even then, techniques exist that allow for an estimation of the ground-state probability~\cite{yoshimura3}.

The locally adiabatic ramp is more complicated to determine~\cite{richerme2}. We start by calculating the excitation spectra $\Delta(B)=E_{1\rm ex}-E_{\rm g.s.}$ for the first coupled excited state relative to the ground state. Then we determine the adiabaticity parameter $\gamma$ from the relation
\begin{equation}
\gamma = \frac{t_f}{\int_0^{B_0}dB\frac{1}{\Delta^2(B)}},
\label{eq: loc_ad}
\end{equation}
where $B_0=5|J_0|$ is the initial magnetic field and $t_f$ is the total experimental time for the ramp. Note that because the initial state corresponds to the ground state for $B\rightarrow\infty$, the locally adiabatic protocol actually starts with a magnetic-field quench. With the adiabaticity parameter determined, 
the magnetic field ramp $B^z(t)$ is found from solving the first-order differential equation
\begin{equation}
\frac{dB^z(t)}{dt}=\frac{1}{\gamma}\Delta^2[B^z(t)].
\label{eq: de}
\end{equation}
After the field ramp profile has been found, we time-evolve the Hamiltonian, with the time-dependent field ramp, by employing the Crank-Nicolson algorithm~\cite{crank_nicolson} choosing a step size that is small enough to guarantee that unitarity is preserved and that the final ground-state probability does not significantly change when the step size is further reduced.

The initial state for both cases is the state where the spins are completely aligned with the field, corresponding to $B^z\gg |J_0|$, but the field ramp always starts with an initial field that is much lower than this (it is equal to $5|J_0|$ for the locally adiabatic ramp and is often much smaller for the bang-bang
shortcut).

\begin{widetext}
\begin{figure*}[t!]
\centering{
\includegraphics[width=1.7in]{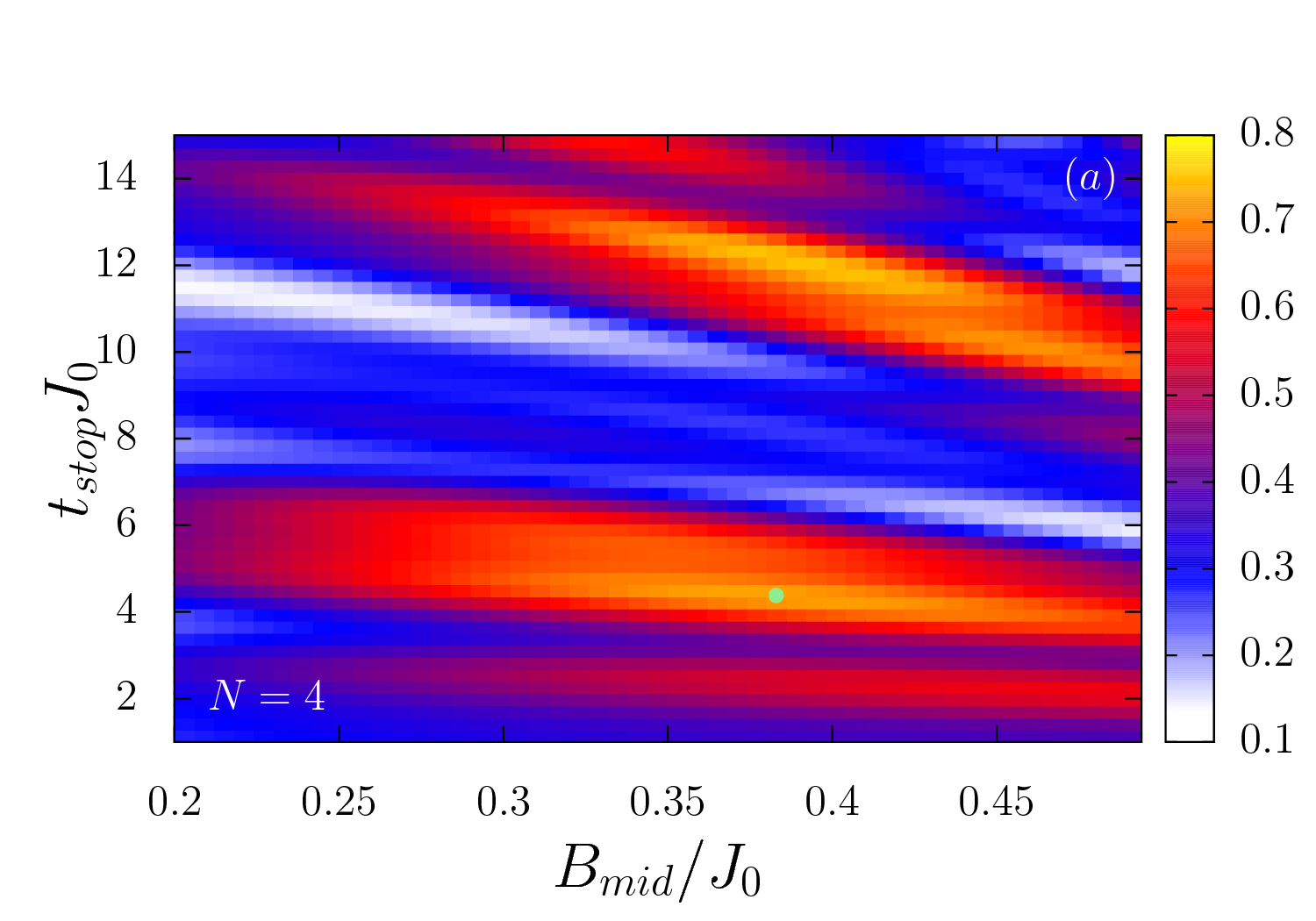}
\includegraphics[width=1.7in]{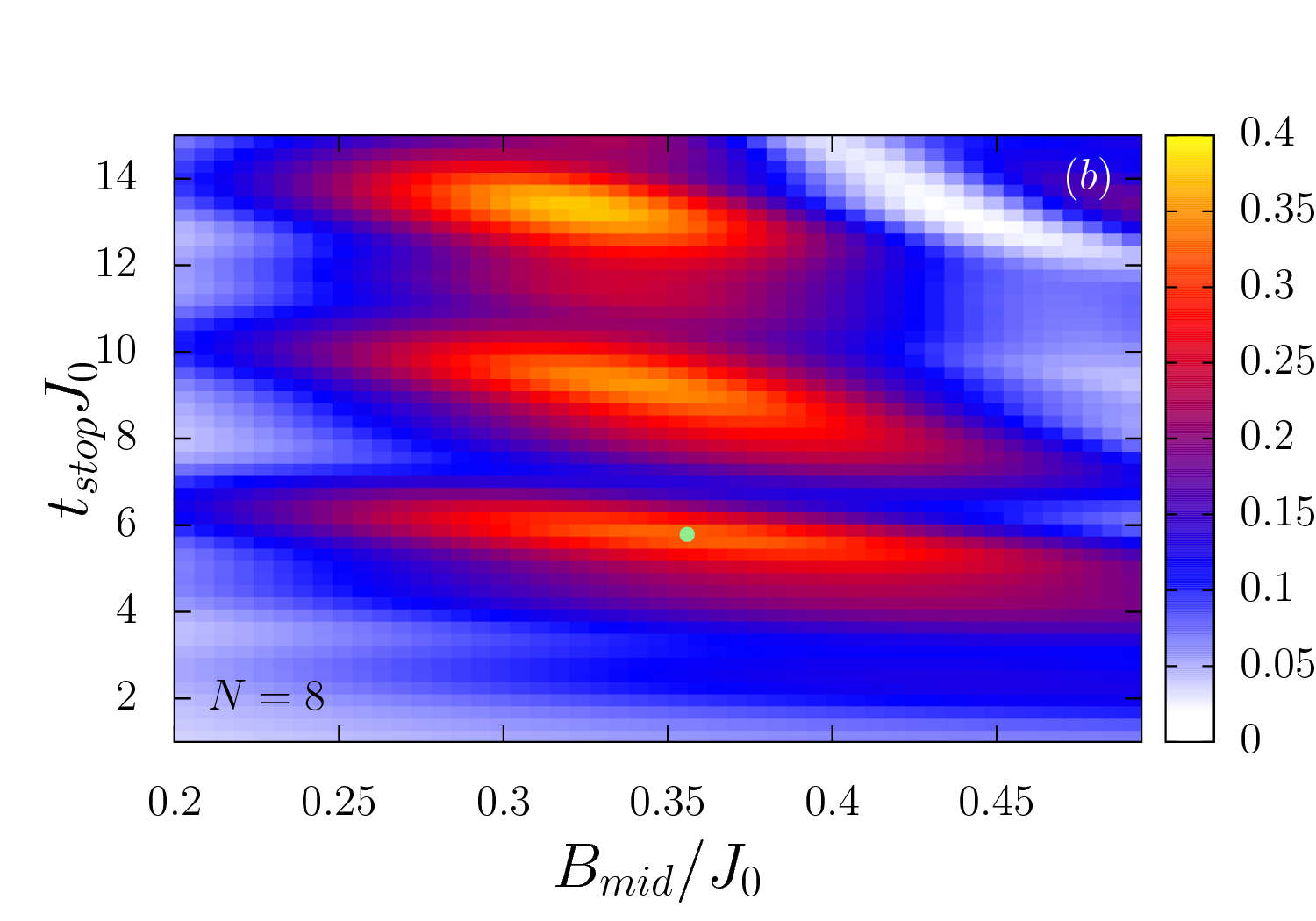}
\includegraphics[width=1.7in]{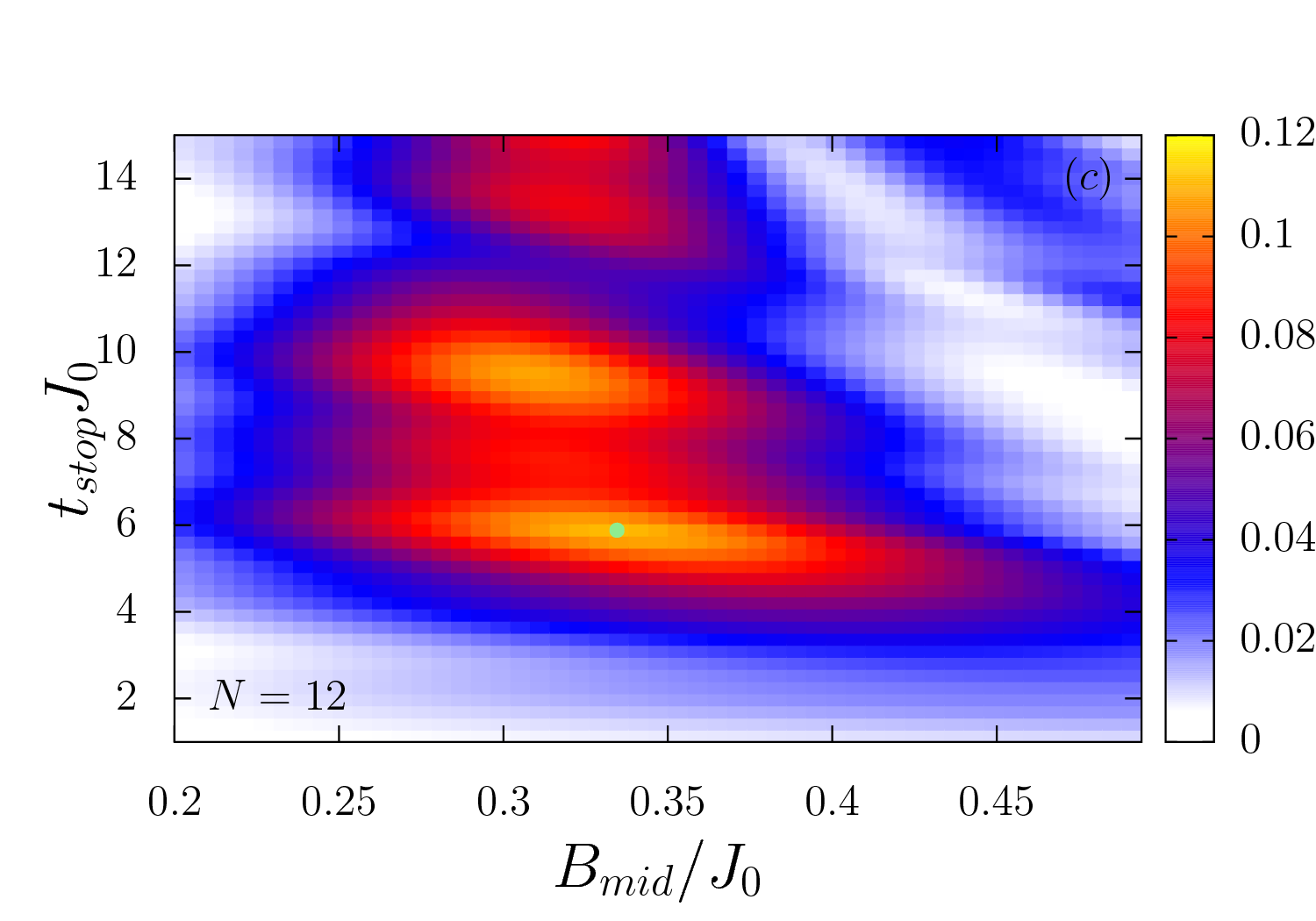}
\includegraphics[width=1.7in]{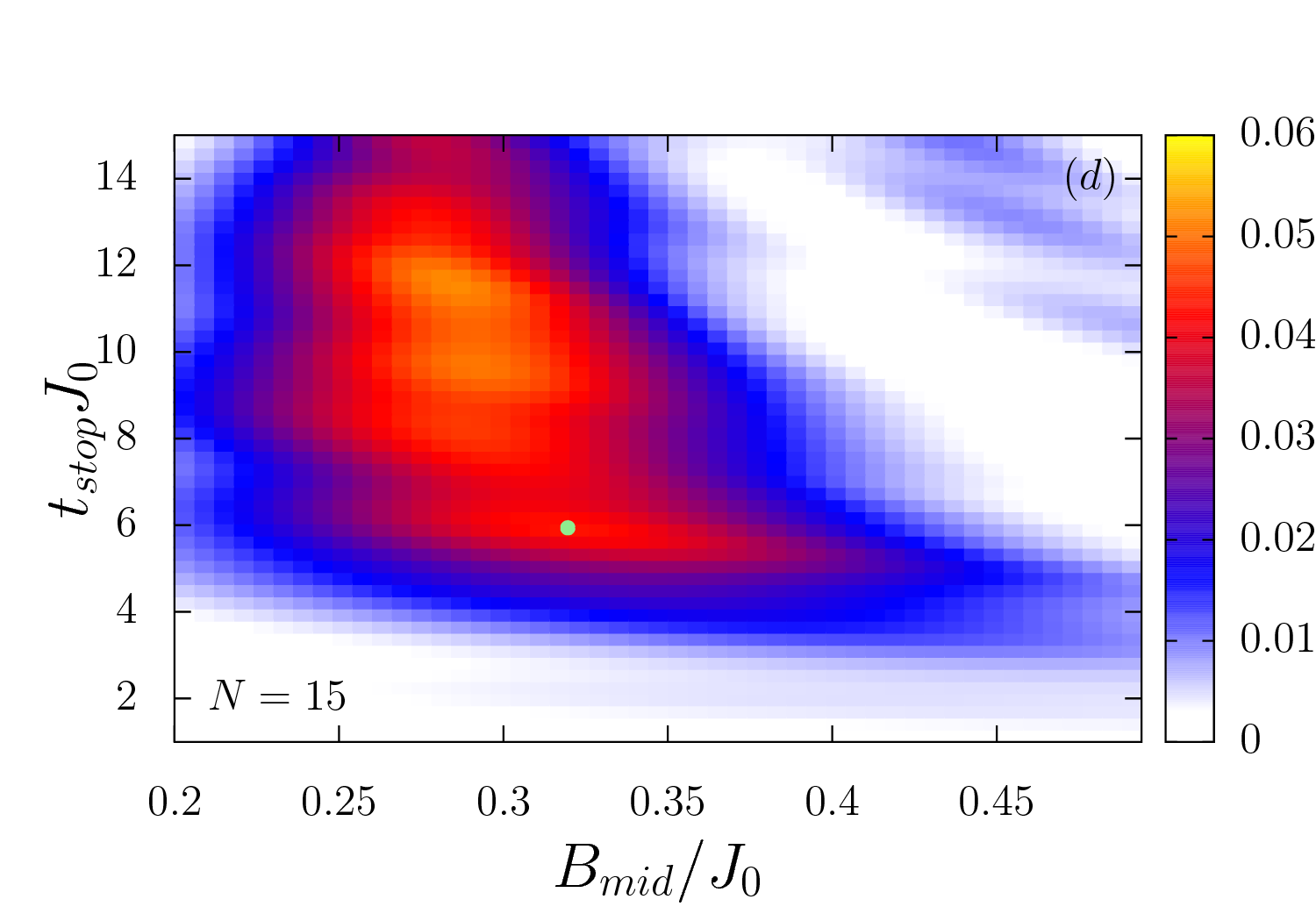}}
\caption{(Color online.) False color plots of the ground-state probability for the bang-bang shortcut to adiabaticity as a function of hold time and quench magnetic field. Different panels correspond to different $N$ values: (a) $N=4$; (b) $N=8$; (c) $N=12$; and (d) $N=15$. The trap parameters are adjusted so that the power law for the decay of the Ising exchange coefficient is approximately $\alpha=1.05$. Note the interesting plateaus that form, and remain at specific times. The light green circle marks the optimized value for the time interval of $t_{\rm exp}< 6$~ms. Note that the false color scale changes in each panel.\label{fig: bang-bang}}
\end{figure*}

\begin{figure*}[t!]
\centering{
\includegraphics[width=1.7in]{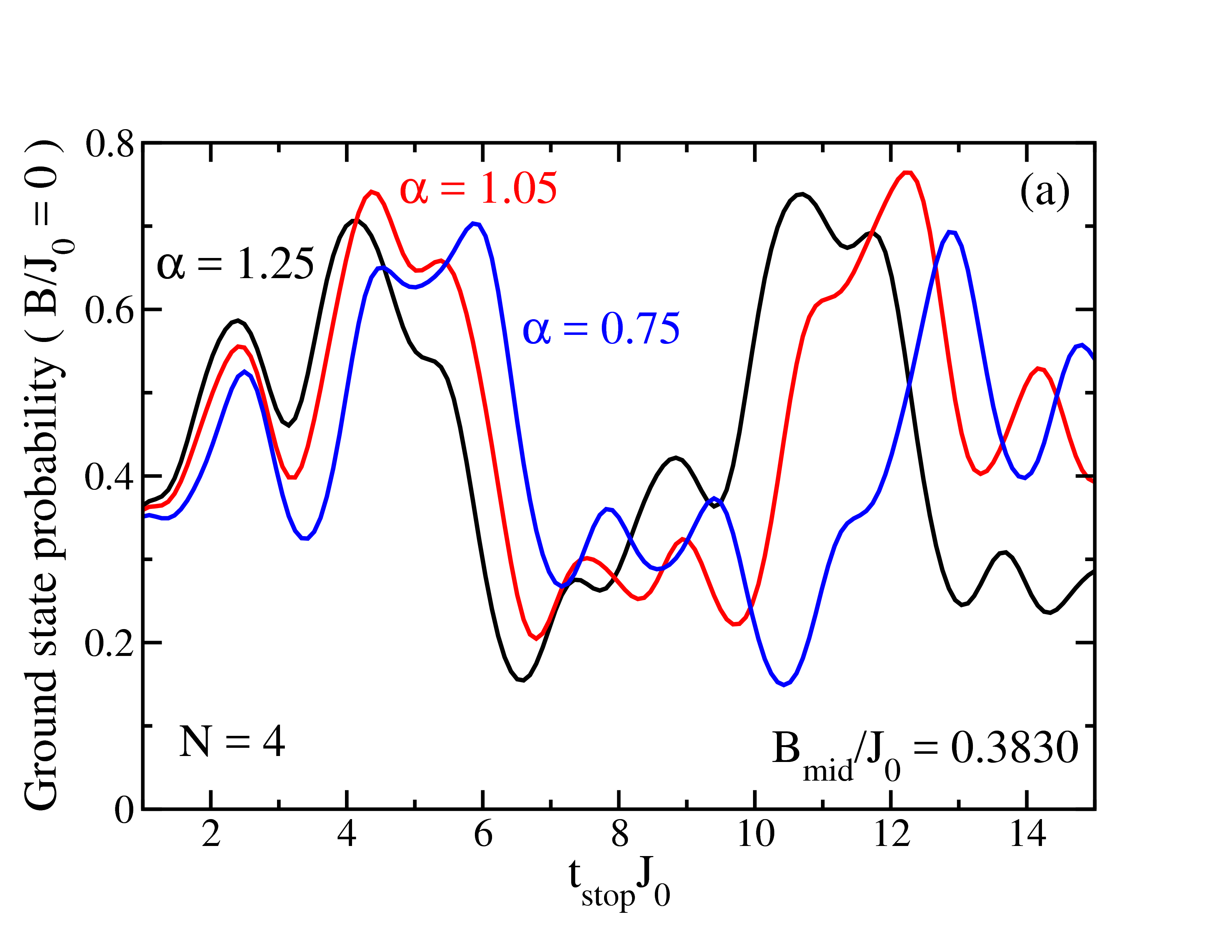}
\includegraphics[width=1.7in]{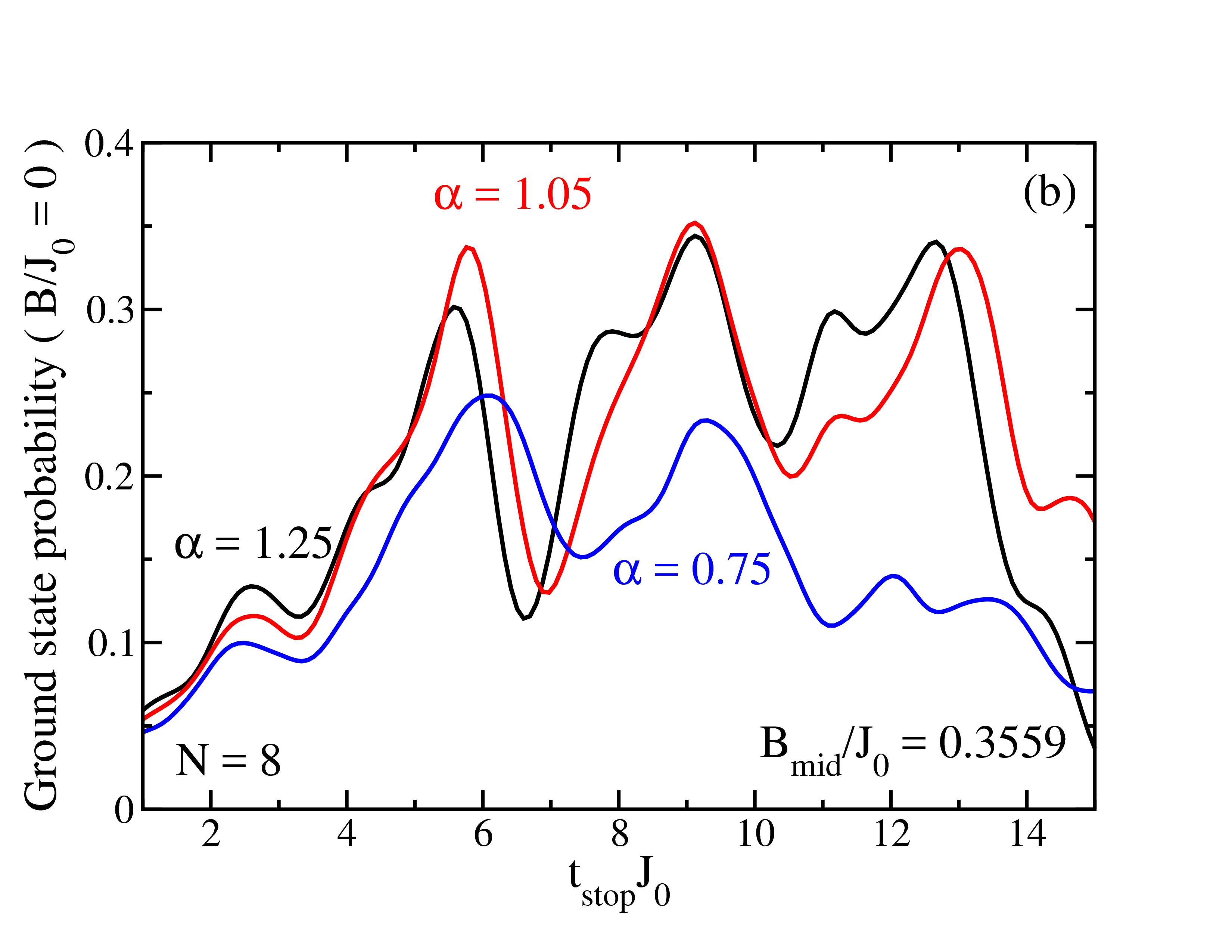}
\includegraphics[width=1.7in]{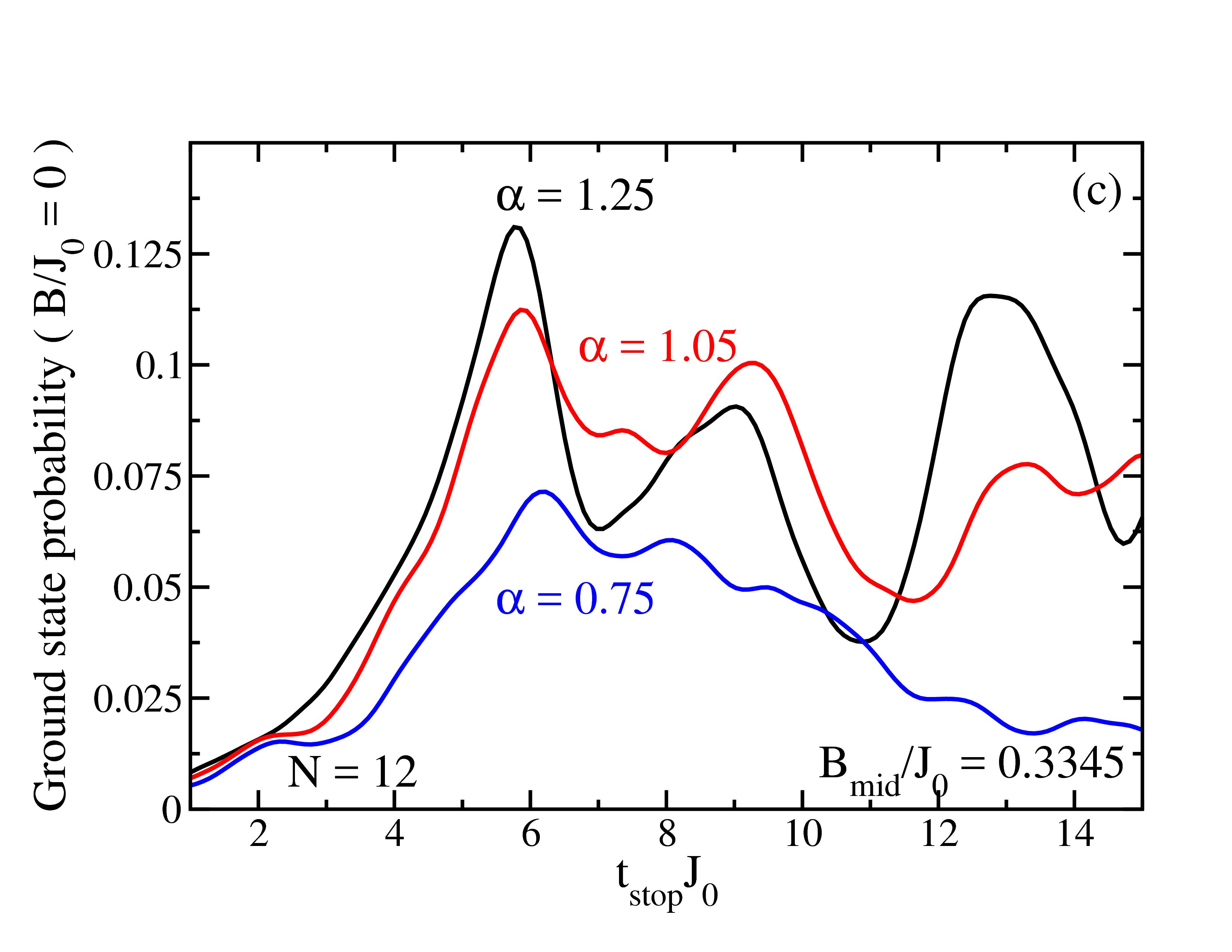}
\includegraphics[width=1.7in]{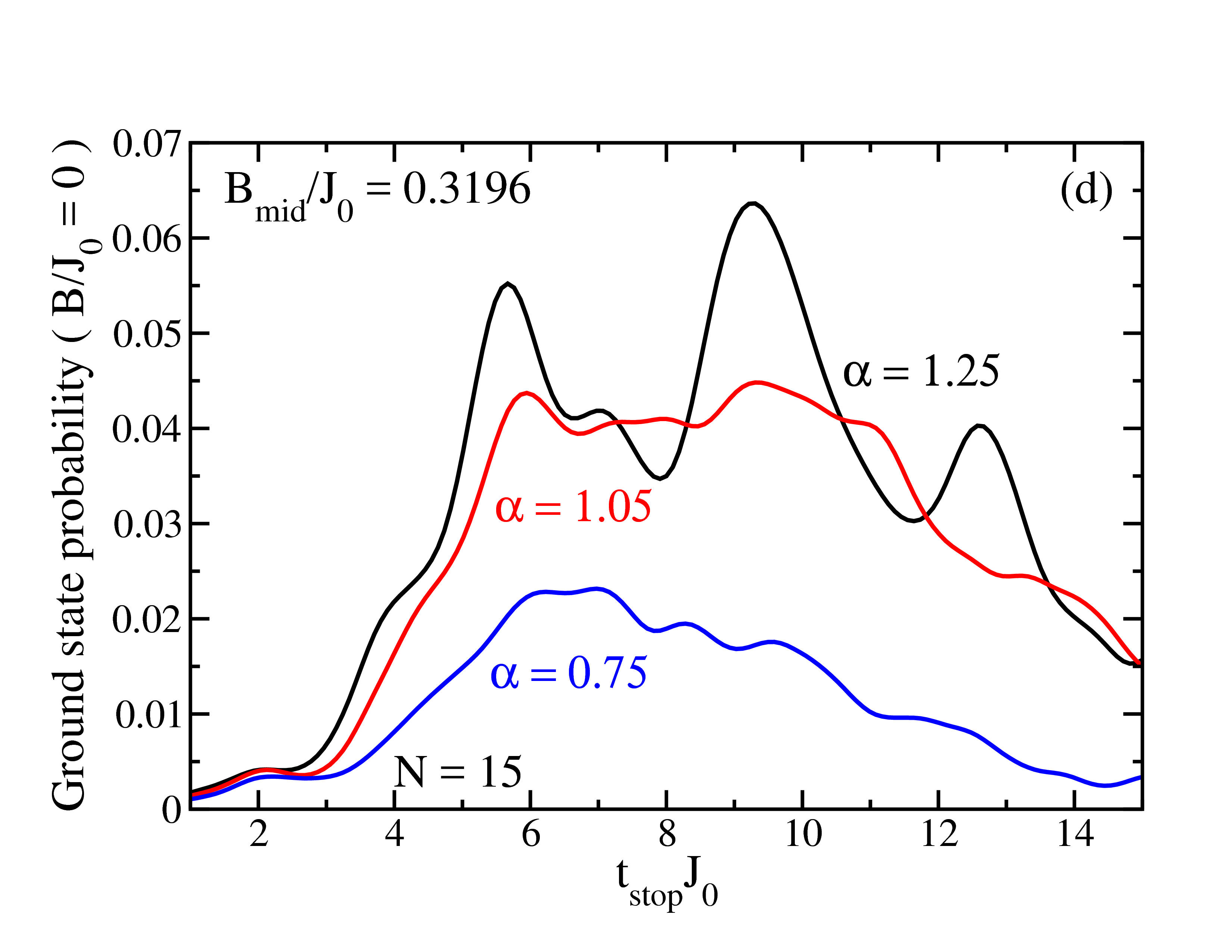}}
\caption{(Color online.)  Vertical cuts through the false-color optimization plots for the bang-bang shortcut with three different power laws for the exchange coefficients: $\alpha = $ 0.75, 1.05, and 1.25.
Different panels correspond to different size ion chains: (a) $N=4$; (b) $N=8$; (c) $N=12$; and (d) $N=15$. \label{fig: bang-bang2}}
\end{figure*}
\end{widetext}

\noindent
{\it Results.} We choose the total experimental run time to be 6~ms. This time is somewhat longer than current experiments (which ran on the order of 2.4~ms~\cite{rajibul2}), but is certainly within reach with available technology. This time is long enough that it allows us to compare the results of the bang-bang shortcut to adiabaticity to the locally adiabatic ramp for chain sizes up to $N=15$. We present only a small selection of our results here that illustrate the most important physical behavior. In Fig.~\ref{fig: bang-bang}, we show false-color images of the probability to be in the final ground-state after the bang-bang shortcut for a given quench field (horizontal axis) and a given hold time
(vertical axis).  Note that there are high probability plateaus (primarily red and orange) and that the plateaus remain over a wide range of varying $N$ in Figs.~\ref{fig: bang-bang}~(a-d). As the system size increases, these plateau are pushed upwards to longer hold times, and the area decreases, but they remain robust for a wide
range of parameters and are the key behind the success of the bang-bang shortcut. This structure indicates that the bang-bang shortcut should be viable for large values of $N$ as long as the experiment can be run for long enough times. The push of the hold time to larger values as $N$ increases must be related to the quantum speed limit~\cite{quantum_speed_limits}, but the precise relationship is not obvious to us.

In Fig.~\ref{fig: bang-bang2}, we plot vertical cuts through the false-color plots that show the final ground-state probability for the bang-bang shortcut at the optimal quench field near 6~ms for different power laws of the spin-spin exchange parameters. One notes immediately that the stability of the plateaus is improved as the power law gets larger, indicating that the shortcut will work better for 
shorter-range spin-spin couplings. One can also clearly see that as $N$ increases, the plateau at smaller hold times disappear, but the ones at higher hold times remain robust. 

\begin{figure}[thb]
\vskip -0.15in
\centering{
\includegraphics[width=0.95\columnwidth]{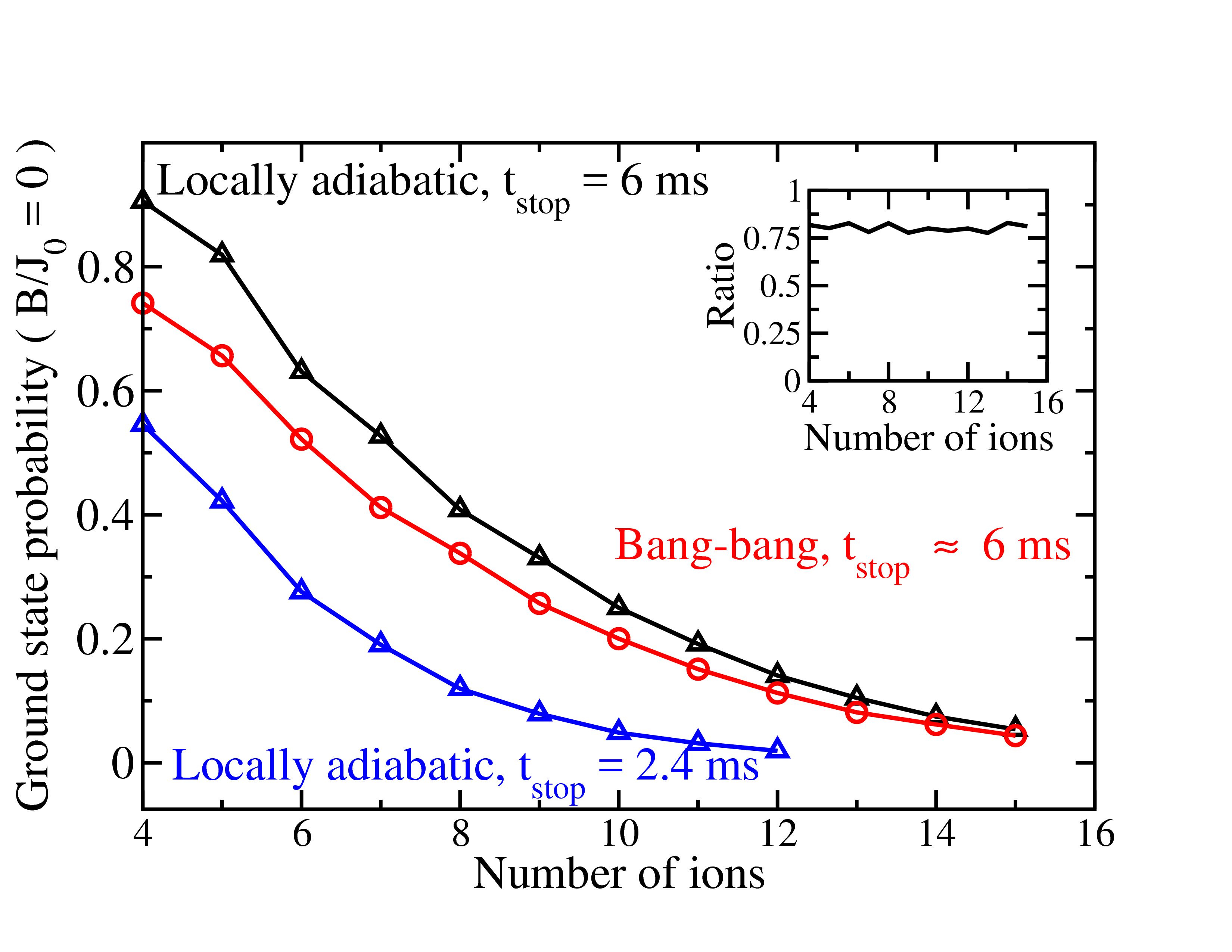}
}
\caption{(Color online.) 
Comparison of the final ground-state probability for the bang-bang shortcut versus the locally adiabatic ramp. The plot is a function of the number of ions in the chain. We also compare to the locally adiabatic results for a shorter time, which is a more typical experimental time at the present. Inset is the ratio of the ground-state percentage for the bang-bang shortcut to the locally adiabatic ramp. One can clearly see that the bang-bang shortcut produces about 80\% of the locally adiabatic probability for the ground state.
\label{fig: compare}
}
\end{figure}

We next compare the final ground-state probability for the two different techniques in Fig.~\ref{fig: compare}; inset, we show the ratio of the results. One can see that the locally adiabatic ramp always does do better than the bang-bang shortcut, which seems to remain robustly at around 80\% of the locally adiabatic ground-state probability (the fluctuations are most likely coming from the fact that the bang-bang optimized hold time is often somewhat different from 6~ms). However, because the locally adiabatic ramp requires so much detailed knowledge of the system being simulated, it is likely to be much more inconvenient to use in practice. One loses very little in terms of the ground-state probability with the bang-bang protocol, and it is dramatically easier to implement, especially for a system with a complex ground state that is not already known. 

\noindent
{\it Conclusions.}
We examined the possibility of using a a bang-bang shortcut to adiabaticity in ion-trap-based quantum simulation as a way to optimize the ground-state probability for adiabatic state preparation.  While we were not able to produce better results than other techniques, the ease of implementing this protocol will likely make it useful within future experiments. We found interesting stable plateaus formed in the plot of the final ground-state probability as a function of the quench field and the hold time. This illustrated not only why the bang-bang approach works but also showed that one needs to go to longer times for larger systems to be able to continue to optimize the ground-state probability. But in general, our results also show that when a system has significant frustration, no technique can maintain a high probability in the ground state, and so it is more useful to consider working with the diabatic distribution of excited states that ensues. In some cases, these distributions can closely mimic thermal distributions~\cite{lim}, but this does not often occur for frustrated spin systems. We examined the distributions of excited states for some of the different systems studied here, and found that the locally adiabatic distribution was not too thermal, but did have a preponderance of the excitations towards the lower energy part of the spectrum. The bang-bang shortcut had an even more athermal distribution, but it actually had lower probabilities of excitation for the low-energy states and modest excitation (almost uniformly distributed) for the higher-energy states (figures not shown here). This indicates that the bang-bang shortcut might have an advantage in preparing specific low-energy excitations, because the other excitations are low in probability. Another advantage might be with regards to phonon creation, especially if the continuous change in time of the Hamiltonian with the locally adiabatic ramp actually creates more phonons \{this is quantitatively determined by the magnitude of $\int dt B^z(t)$~\cite{wang}\}. But that would have to be part of a different study on this topic.

We hope that experimentalists will consider employing quantum control ideas like the bang-bang shortcut within their experiments in the near future as they can be used to gain an even better understanding of how these quantum simulators work.

\acknowledgements

We acknowledge useful discussion with Adolfo del Campo.
This work was supported by the National Science Foundation under grant number PHY-1314295.  JKF was also supported by the McDevitt bequest at Georgetown University. BTY was also supported from the
Achievement Rewards for College Scientists Foundation.

\end{document}